\documentstyle[12pt]{article}

\begin{document}

\baselineskip=18.5pt plus 0.2pt minus 0.1pt
\parskip = 6pt plus 2pt minus 1pt

\catcode`\@=11

\newif\iffigureexists
\newif\ifepsfloaded
\openin 1 epsf.sty
\ifeof 1 \epsfloadedfalse \else \epsfloadedtrue \fi
\closein 1
\ifepsfloaded
    \input epsf.sty
\else
    \immediate\write20{>Warning:
         No epsf.sty --- cannot embed Figures!!}
\fi
\def\checkex#1 {\relax
    \ifepsfloaded \openin 1 #1
        \ifeof 1 \figureexistsfalse
        \else \figureexiststrue
        \fi \closein 1
    \else \figureexistsfalse
    \fi }

\def\epsfhako#1#2#3#4#5#6{
\checkex{#1}
\iffigureexists
    \begin{figure}[#2]
    \epsfxsize=#3
    \centerline{\epsffile{#1}}
    {#6}
    \caption{#4}
    \label{#5}
    \end{figure}
\else
    \begin{figure}[#2]
    \caption{#4}
    \label{#5}
    \end{figure}
    \immediate\write20{>Warning:
         Cannot embed a Figure (#1)!!}
\fi
}

\ifepsfloaded
\checkex{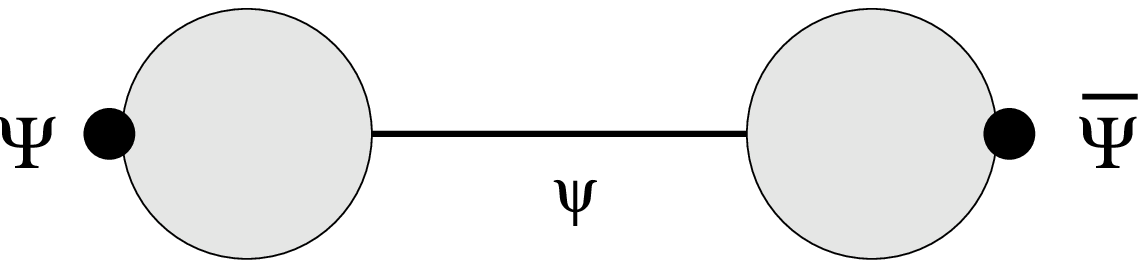}
    \iffigureexists \else
    \immediate\write20{>EPS files for Fig. 1 is packed
     in a uuecoded compressed tar file}
    \immediate\write20{>appended to this LaTeX file.}
    \immediate\write20{>You should unpack them and LaTeX again!!}
    \fi
\fi


\renewcommand{\thefootnote}{\fnsymbol{footnote}}
\renewcommand{\theequation}{\thesection.\arabic{equation}}
\renewcommand{\ne}{\nonumber}

\newcommand{\reseteqnum}{\setcounter{equation}{0}}
\newcommand{\Fmn}{\frac{1}{2g_{\rm YM}^2}{\rm tr}F_{\mu\nu}^2}
\newcommand{\dB}{\delta_{\rm B}}
\newcommand{\dA}{\delta_{\rm A}}
\newcommand{\edB}{\widetilde{\delta}_{\rm B}}
\newcommand{\edA}{\widetilde{\delta}_{\rm A}}
\newcommand{\qB}{Q_{\rm B}}
\newcommand{\eqB}{\widetilde Q_{\rm B}}
\newcommand{\p}{\partial}
\newcommand{\D}{{\cal D}}
\newcommand{\bgamma}{\overline{\gamma}}
\newcommand{\bc}{\overline{c}}
\newcommand{\bPsi}{\overline{\Psi}}
\newcommand{\bpsi}{\overline{\psi}}
\newcommand{\bB}{\overline{B}}
\newcommand{\bbeta}{\overline{\beta}}
\newcommand{\Int}{\int\!}
\newcommand{\tr}{{\rm tr}}
\newcommand{\gYM}{g_{\rm YM}}
\newcommand{\OSp}{OSp(4/2)}
\newcommand{\vev}[1]{\langle{\rm T} #1\rangle}
\newcommand{\VEV}[1]{\left\langle{\rm T} #1\right\rangle}
\newcommand{\Seff}{S_{\rm eff}}
\newcommand{\bm}[1]{\mbox{\boldmath $#1$}}
\newcommand{\SUNR}{SU$(N)_{\rm R}$}
\newcommand{\SUNL}{SU$(N)_{\rm L}$}
\newcommand{\pslash}{p\kern-1ex /}
\newcommand{\auxfields}{$(a_\mu,\gamma,\bgamma,\beta,\psi)$}
\newcommand{\PGMfields}{$(g,c,\bc,B)$}
\newcommand{\CD}{{D}}
\newcommand{\qD}{{\cal D}}
\newcommand{\Dslash}{{\cal D}\kern-1.5ex /}
\newcommand{\Bk}{{B_k}}

\catcode`\@=12

\begin{titlepage}
  \title{ \hfill
\parbox{4cm}{\normalsize KUNS-1325\\HE(TH)~95/04\\hep-th/9502083}\\
\vspace{1cm}
Color Confinement in Perturbation Theory\\
from a Topological Model
}
\author{Hiroyuki Hata\thanks{e-mail address: \tt
    hata@gauge.scphys.kyoto-u.ac.jp}
  {\,}\thanks{Supported in part by Grant-in-Aid for Scientific
    Research from Ministry of Education, Science and Culture (\#
    06640390).} {} and Yusuke Taniguchi\thanks{e-mail address: \tt
    tanigchi@gauge.scphys.kyoto-u.ac.jp} {\,}\thanks{JSPS Research
    Fellow.  Supported in part by Grant-in-Aid for Scientific Research
    from Ministry of Education, Science and Culture (\# 3113).}
\\
{\normalsize\em Department of Physics, Kyoto University}\\
{\normalsize\em Kyoto 606-01, Japan}}
\date{\normalsize February, 1995}
\maketitle \thispagestyle{empty}

\begin{abstract}
  \normalsize

Color confinement by the mechanism of Kugo and Ojima can treat
confinement of any quantized color carrying fields including dynamical
quarks.
However, the non-perturbative condition for this confinement has been
known to be satisfied only in the pure-gauge model (PGM), which is
a topological model without physical degrees of freedom.
Here we analyze the Yang-Mills theory by adding physical
degrees of freedom as perturbation to PGM. We find that quarks and
gluons are indeed confined in this perturbation theory.

\end{abstract}
\end{titlepage}

\section{Introduction}

The understanding of the color confinement mechanism is one of the
most fascinating problems in QCD.
In this paper we shall study the color confinement mechanism proposed
by Kugo and Ojima (KO) \cite{KO}. It is based on the BRST quantized
Yang-Mills theory (coupled to the quark field $\Psi$) described by the
Lagrangian,
\begin{equation}
{\cal L} = \Fmn(A) +
\bPsi (i \Dslash(A) -m) \Psi -i \dB G(A_\mu, c,\bc, B),
\label{eq:YM}
\end{equation}
with $F_{\mu \nu}(A)\equiv\p_\mu A_\nu - \p_\nu A_\mu + [A_\mu,A_\nu]$
and $\qD_\mu(A) \equiv \p_\mu + A_\mu$.\footnote{
We restrict the gauge group to SU$(N)$.
The field variables $\phi=A_\mu,c,\bc,B$ are Lie algebra valued
and are expressed as $\phi=\sum_{a=1}^{N^2-1}\phi^a t^a$ in terms of
Hermitian fields $\phi^a$ and the (anti-Hermitian) basis $t^a$ with
the normalization $\tr(t^at^b)=-(1/2)\delta^{ab}$.}
In Eq.\ (\ref{eq:YM}), $\dB$ is the BRST transformation, and
$G(A_\mu,c,\bc,B)$ specifies the gauge fixing.
In this paper we adopt as $G$ either
$G_\alpha=\tr\left[\bc\left(\p^\mu A_\mu -\alpha B\right)\right]$
corresponding to the Feynman type gauge or
\begin{equation}
G_{\rm OSp}=\frac{2}{\lambda}\dA\!\left\{\tr\left(A_\mu^2 +
2ic\bc\right)\right\} ,
\label{eq:GOSp}
\end{equation}
of the \OSp\ symmetric gauge \cite{OSp} ($\lambda$ is the gauge
parameter and $\dA$ is the anti-BRST transformation \cite{dA} defined
in the Appendix).

The confinement by the KO mechanism is in the sense that
there are no {\em physical color-carrying} asymptotic fields.
In contrast to the Wilson loop criterion \cite{Wilson}, the KO
mechanism treats confinement of {\em quantized\/} quarks and
gluons.
According to Kugo and Ojima \cite{KO}, a sufficient
condition for this confinement to be realized is that the
BRST-exact conserved color current of the system (\ref{eq:YM}),
\begin{equation}
N_\mu=-i\dB\Bigl(D_\mu\bc\Bigr)
=-i\dB\Bigl(\p_\mu\bc+[A_\mu,\bc]\Bigr) ,
\label{eq:N}
\end{equation}
contains no (Nambu-Goldstone-like) massless one-particle mode.
If this condition is satisfied, the color charge $Q^a$ is written in a
well-defined manner as
$Q^a=\left\{\qB, \int d^3x \left(D_0\bc\right)^a\right\}$.
The BRST-exact expression $Q^a=\left\{\qB,*\right\}$ ensures that any
color non-singlet asymptotic states are necessarily BRST unphysical
and hence unobservable.

Differently to the Wilson loop criterion, the KO mechanism
seems to have little relationship to the observable QCD dynamics
creating mesons and baryons.
Due to this property, however, the KO confinement mechanism may be
studied rather simply without being bothered by complicated observable
dynamics. An extreme example of the realization of this expectation is
the pure-gauge model (PGM) \cite{PGM-H,PGM-HK}. This is a toy model
for the KO confinement mechanism.
The PGM is obtained from the Yang-Mills system (\ref{eq:YM}) by
restricting the gauge field $A_\mu$ to the pure-gauge configuration,
$A_\mu=g^\dagger\p_\mu g$, and therefore is described by the
``topological'' Lagrangian:
\begin{equation}
{\cal L}_{\rm PGM} = -i \dB G(g^{\dagger}\partial_\mu g,c,\bc,B).
\label{eq:PGM}
\end{equation}
Although the PGM has only unphysical gauge-modes, the KO condition
that $N_\mu$ is free from the massless mode is still a non-trivial
problem of dynamics. If we adopt the \OSp\ symmetric gauge with
$G_{\rm  OSp}$ (\ref{eq:GOSp}), the PGM in four dimensions becomes
equivalent to the chiral model in two dimensions owing to the
Parisi-Sourlas mechanism \cite{ParisiSourlas}.
This equivalence and the fact that the chiral model in two dimensions
is realized in the disordered phase with a mass gap
\cite{PolyakovWiegmann} implies that the KO condition is actually
satisfied in the PGM with the \OSp\ symmetric gauge \cite{PGM-HK}.
This property is expected to be shared with the ordinary Feynman
type gauge \cite{PGM-H}.
A lesson we learn from PGM is that, among the degrees of freedom in
$A_\mu$, the large fluctuation of the gauge-mode $g(x)$, namely the
mode in the direction of the gauge transformation, is important for
the KO confinement mechanism (cf.\ Ref.\ \cite{Hata-restoration}).

The purpose of this paper is to present a method to perturbatively
introduce physical degrees of freedom into the PGM and study the KO
confinement mechanism in the real Yang-Mills theory.
Roughly speaking what we would like to do is to express the gauge
field as $A_\mu= g^\dagger\p_\mu g + \mbox{(physical-modes)}$
and treat the physical-modes as perturbation to PGM.
The parameter of our perturbation expansion is simply the gauge coupling
constant $\gYM$ (note that the PGM is obtained from the Yang-Mills
system (\ref{eq:YM}) in the vanishing coupling constant
limit $\gYM\to 0$).
We see that, in this perturbation method, quark and gluon fields are
confined by the KO mechanism due to the large fluctuation of the
gauge-mode $g(x)$.
A similar attempt to adding physical modes to the PGM was made by
Izawa \cite{Izawa} using the BF formulation.

The organization of the rest of this paper is as follows.
In Sec.\ \ref{sec:perturbation}, we present the general framework of
our perturbation expansion from PGM, and in Sec.\ \ref{sec:KO}, we
apply it to the study of the KO confinement mechanism.
In Sec.\ \ref{sec:FT}, we generalize our formulation to the finite
temperature case.
The final section (Sec.\ \ref{sec:discussion}) is devoted to
discussion on the remaining problems.
In the Appendix, we summarize the (anti-)BRST transformations.

\reseteqnum
\section{Perturbation expansion around the PGM}\label{sec:perturbation}

For carrying out the perturbation expansion around the PGM, we shall
first modify the Yang-Mills system (\ref{eq:YM}) by introducing the
new auxiliary fields.
That is, we rewrite the system in such a way that the gauge
field is expressed as a sum of the pure-gauge mode
$g^{\dagger} \partial_\mu g$ and the other (physical) modes.
Consider the partition function
\begin{eqnarray}
Z[J]=\Int  \D A_\mu \D c \D \bc \D\! B  \D\Psi \D\bPsi
\exp\left\{i\Int dx \Bigl({\cal L} + J\!\cdot\!\Phi\Bigr)\right\},
\label{eq:Z1}
\end{eqnarray}
where ${\cal L}$ is given by (\ref{eq:YM}) and
$J\cdot\Phi\equiv \tr\left(J^\mu A_\mu + J_c c + J_{\bc}\bc + J_B
B\right)+ \overline{j}\Psi + j\bPsi$ is the source term.
Our strategy is to carry out the Faddeev-Popov (FP) trick
\cite{FPtrick} again to (\ref{eq:Z1}).
We define the FP determinant $\Delta[A]$ as usual by
\begin{equation}
  1 = \Delta [A] \Int {\cal D}g \prod_{x} \delta \left( \p^\mu
  A^{g^{-1}}_\mu (x) \right),
\label{eq:FPD}
\end{equation}
where $g(x)$ is an element of the gauge group SU$(N)$ and
$A_\mu^g\equiv g^\dagger\left(\p_\mu + A_\mu\right)g$.
We have adopted the Landau gauge for (\ref{eq:FPD}).
We multiply (\ref{eq:FPD}) to the partition function
(\ref{eq:Z1}) and express the fields $A_\mu$ and $\Psi$
as the gauge transform of the new fields $a_\mu$ and $\psi$:
\begin{eqnarray}
A_\mu \!&=&\! a^g_\mu \equiv g^{\dagger} \p_\mu g
+ g^{\dagger} a_\mu g,\ne\\
\Psi \!&=&\! \psi^{g} \equiv g^{\dagger} \psi .
\label{eq:expansion}
\end{eqnarray}
Using the gauge invariance of the FP determinant
$\Delta [A] = \Delta [A^g]$ and expressing the product
$\Delta [a] \prod \delta \left( \p^\mu a_\mu \left(x\right) \right)$
in terms of the integration over the new (anti-)ghosts $\gamma$,
$\bgamma$ and the multiplier field $\beta$,
\begin{eqnarray}
\Delta [a] \prod_{x} \delta \left( \p^\mu a_\mu \left( x \right)
\right) \!&=&\! \Int \D \gamma \D \bgamma \D \beta
\exp\left\{ i \Int
dx \left( \beta^a \p^\mu a^a_\mu + i\bgamma^a \p^\mu D_\mu(a)
\gamma^a \right)\right\}\ne\\
\!&=&\! \Int \D \gamma \D \bgamma \D \beta \exp\left\{ i
\Int dx \left( -i \edB H\left(a_\mu, \gamma, \bgamma, \beta\right)
\right)\right\}, \label{eq:FPG}
\end{eqnarray}
the partition function (\ref{eq:Z1}) is rewritten as
\begin{eqnarray}
Z[J] &=& \Int \D g \D c \D \bc \D B\Int\D a_\mu \D
\gamma \D   \bgamma \D \beta\D \bpsi \D \psi \ne\\
&\times&\! \exp\Biggl\{ i\Int dx
\biggl[ \Fmn(a) +\bpsi
\left( i\Dslash(a) -m\right) \psi -i \edB
H\left( a_\mu, \gamma, \bgamma, \beta\right) \ne\\
&&-i \dB  G\left( g^{\dagger}\left( a_\mu + \p_\mu \right) g, c, \bc,
B\right) + J\!\cdot\!\Phi\biggr]\Biggr\} .
\label{eq:ZF}
\end{eqnarray}
In Eqs.\ (\ref{eq:FPG}) and (\ref{eq:ZF}), we have
$H=\tr\left(\bgamma\, \p^\mu a_\mu\right)$, and the new BRST
transformation $\edB$ is given in the Appendix.
$A_\mu$ and $\Psi$ in the source term $J\cdot\Phi$ should be
understood to be expressed in terms of $g$, $a_\mu$ and $\psi$ as
given in Eq.\ (\ref{eq:expansion}).

Eq.\ (\ref{eq:ZF}) combined with Eq.\ (\ref{eq:expansion}) is the
basic formula for our perturbation theory.
The original gauge field $A_\mu$ is expressed in terms of the new
gauge field $a_\mu$ in Landau gauge and the gauge fluctuation $g$
around $a_\mu$.
The path-integrations in (\ref{eq:ZF}) consist of those over the PGM
fields $(g,c,\bc,B)$ and those over the new fields
$(a_\mu,\gamma,\bgamma,\beta,\psi)$.
Corresponding to two kinds of the BRST transformations, $\dB$ and
$\edB$, the system (\ref{eq:ZF}) has \SUNR\ and \SUNL\ global
symmetries; $g(x)\to h_{\rm L}^\dagger g(x)h_{\rm R}$.
The original fields $(A_\mu,c,\bc,B,\Psi)$ and the new ones
\auxfields\ are  transformed by \SUNR\ and \SUNL, respectively.
The color rotation of the original system (\ref{eq:YM}) is \SUNR.

Our formula (\ref{eq:ZF}) has no advantage over the original $Z$
(\ref{eq:Z1}) for calculating gauge invariant quantities.
However, in the study of the KO mechanism, we are interested in the
gauge-variant quantities such as the Green's function
$\VEV{N_\mu A_\nu}$ \cite{KO}, and our formula will prove very useful
as we shall see below.

Now, we shall explain our perturbation expansion around the PGM on
the basis of the formulas (\ref{eq:expansion}) and (\ref{eq:ZF}).
In the following we are interested in the Green's functions of the
original fields $(A_\mu=a_\mu^g,c,\bc,B,\Psi=\psi^g)$ obtained by
differentiating $Z[J]$ (\ref{eq:ZF}) with respect to $J$.
Our expansion around the PGM is to carry out the integration over the
auxiliary fields \auxfields\ in (\ref{eq:ZF}) using perturbation
expansion in powers of the coupling constant $\gYM^2$.
The remaining integration in (\ref{eq:ZF}) over the PGM fields
\PGMfields\ should be treated non-perturbatively.

Let us consider this expansion more concretely.
First, it should be noticed that in the exponent of (\ref{eq:ZF}) the
interactions between the PGM fields $(g,c,\bc,B)$ and the auxiliary
ones $(a_\mu,\gamma,\bgamma,\beta,\psi)$ come from
1) the gauge fixing term $-i\dB G$,
2) the $g^\dagger a_\mu g$ term in $A_\mu=a_\mu^g$ of the
source term $J^\mu A_\mu$, and
3) $\Psi=g^\dagger\psi$ in the source term $\bar j\Psi + j\bPsi$.
Especially the first contribution 1) is as follows. Taking
the \OSp\ symmetric gauge, we have
\begin{eqnarray}
  -i \dB G_{\rm OSp}\kern-1cm&& \left(
  g^{\dagger}\left(a_\mu + \p_\mu\right)g, c, \bc, B\right)
=\frac{2i}{\lambda} \dA \dB \ \tr \left[ (g^{\dagger}
a_\mu g + g^{\dagger} \p_\mu g)^2 +2i c\bc \right]\ne\\
&=&\!\! -i \dB G_{\rm OSp}\left(
g^{\dagger}\p_\mu g,  c, \bc, B\right)
-\frac{4i}{\lambda} \tr\left[
 a_\mu\,\dA \dB (g \p^\mu g^{\dagger}) \right] , \label{eq:interaction}
\end{eqnarray}
where use has been made of the fact that $a_\mu$ is invariant under
both $\dA$ and $\dB$.
In the last expression of (\ref{eq:interaction}) the first term is
nothing but the PGM Lagrangian (\ref{eq:PGM}), and the second term is
the interaction between the PGM fields and the auxiliary one $a_\mu$.
The separation into the PGM Lagrangian and the
interaction of the form
$\tr\left[a_\mu \dA \dB (g \p^\mu g^{\dagger})\right]$ is also
the case if we take the ordinary Feynman type gauge.

Therefore, in the case $j=\bar j=0$ (i.e., no external quarks), the
generating functional (\ref{eq:ZF}) is rewritten as follows:
\begin{eqnarray}
Z[J]\! &=&\! \Int\D g\D c\D\bc\D B\exp\Biggl\{i\Int dx\biggl[
-i\dB G(g^\dagger\p_\mu g,c,\bc,B) \ne\\
&&+ \tr\left( J_\mu\, g^\dagger\p_\mu g
+ J_c c + J_{\bc}\bc + J_B B\right)\biggr]
+ iW\left[gJ_\mu g^\dagger
-\frac{4i}{\lambda}\dA\dB\left(g\p_\mu g^\dagger\right)\right]
\Biggr\},
\label{eq:Zpert}
\end{eqnarray}
where $W[j_\mu]$ is the generating functional of the connected Green's
function of $a_\mu$ in the auxiliary field sector:
\begin{eqnarray}
\exp iW[j_\mu]=\Int\D a_\mu\D\gamma\D\bgamma\D\beta\D\bpsi\D\psi
\exp\Biggl\{ i\Int dx
\biggl[ \Fmn(a) +\bpsi
\left( i\Dslash(a) -m\right) \psi\ne\\
-i \edB H\left(a_\mu,\gamma,\bgamma,\beta\right) +
\tr j_\mu a_\mu\biggr]\Biggr\} .
\label{eq:W}
\end{eqnarray}
$W[j_\mu]$ should be calculated using ordinary perturbation theory
with coupling constant $\gYM^2$: the diagram contributing to
$W[j_\mu]$ with $N$ external $a_\mu$ and $L$ auxiliary field loops is
multiplied by $\left(\gYM^2\right)^{N+L-1}$.
Explicitly we have
\begin{eqnarray}
W[j_\mu]=\frac{1}{2}\gYM^2\,\tr\!\Int dx\!
\Int dy j_\mu(x)D^{\mu\nu}(x-y)j_\nu(y)
+ O\left(\gYM^4\right),
\label{eq:Wfirst}
\end{eqnarray}
where $D_{\mu\nu}(x)$ is the free propagator in the Landau gauge:
\begin{equation}
D_{\mu \nu}(x)=\Int \frac{d^4 p}{(2\pi)^4}
\frac{1}{p^2-i\varepsilon} \Bigl( g_{\mu \nu}-\frac{p_\mu p_\nu}{p^2}
 \Bigr) e^{ip\cdot x} .
\label{eq:Pa}
\end{equation}
In particular, from Eqs.\ (\ref{eq:Zpert}) and (\ref{eq:Wfirst}) we
reconfirm that the Yang-Mills theory is reduced to the PGM in the
limit $\gYM\to 0$.
The case of the Green's functions containing the quark fields $\Psi$
and $\bPsi$ can be treated similarly.

\reseteqnum
\section{Color confinement}\label{sec:KO}

Now let us consider how the color confinement by the KO mechanism is
realized in the perturbation expansion explained in the last section.
Although the KO confinement condition is satisfied in PGM,
there is no physical modes to be confined.
Differently to the PGM the present model does contain physical degrees
of freedom, and we shall see that they are indeed confined owing to
the disorder of the gauge-mode $g(x)$.
Before starting the discussion, we here stress that we are
interested in the confinement of the \SUNR\ color charge carried by
the original fields $(A_\mu, \Psi)$ and abandon the confinement of
\SUNL\ charge of $(a_\mu, \psi)$.
These two color charges are physically equivalent and both should be
confined in a more complete treatment of Yang-Mills theory.
However, our perturbation expansion discriminates between these two
color charges.

We shall discuss the confinement of \SUNR\ color in two different
ways: one is to see directly the absence of the physical
asymptotic states of $\Psi$ and $A_\mu$, and the other is to study the
sufficient condition for confinement, i.e.,
the absence of the Nambu-Goldstone mode coupled to the BRST-exact
\SUNR\ color current $N_\mu$ (\ref{eq:N}).

First, let us consider the asymptotic fields of quarks and gluons.
The question is whether the two-point functions
$\vev{\Psi \bPsi}$ and $\vev{A_\mu A_\nu}$ (in momentum space) have
discrete poles corresponding to the physical asymptotic states.
However, these discrete poles are absent in the lowest order
of our perturbation expansion. This is seen as follows. Using the
expression (\ref{eq:expansion}) for the original quark field $\Psi$ in
terms of $\psi$ and $g$, we have
\begin{eqnarray}
\VEV{\Psi_i(x)\bPsi_j(0)}=\VEV{(g^\dagger)_{ik}(x)g_{lj}(0)}_{\rm PGM}
\VEV{\psi_k(x) \bpsi_l(0)}_0 + O(\gYM^2),
\label{eq:PsiPsi}
\end{eqnarray}
where $\vev{\cdots}_{\rm PGM}$ denotes the Green's function in the
PGM, and $\vev{\psi_k \bpsi_l}_0=\delta_{kl}/(\pslash -m)$ is the free
propagator ($i,j,\ldots$ are the SU$(N)$ color indices in the
fundamental representation). The point is that the PGM is in the
disordered phase where \SUNL$\otimes$\SUNR\ symmetry is realized
linearly and the SU$(N)$-valued field $g(x)$ has $N^2$ independent
excitation modes;
$\VEV{(g^\dagger)_{ik}g_{lj}}_{\rm PGM}\sim
\delta_{ij}\delta_{kl}/(p^2-M^2)$
with $M^2$ being the mass gap of the PGM
(in particular, we have $\langle g\rangle_{\rm PGM}=0$).
Namely, $\Psi=g^\dagger\psi$ is a genuine two-body composite operator
and there is no interaction between the two in the lowest order in our
expansion.
Therefore, although the auxiliary quark field $\psi$ has an asymptotic
field, the original quark field $\Psi$ does not.
In a word,
the large fluctuation of the gauge-mode $g$ screens the asymptotic
field of the auxiliary field $\psi$.
The KO confinement mechanism requires only that the quark asymptotic
fields, if they exist, are BRST unphysical \cite{KO}. In the present
case, the asymptotic field of $\Psi$ is totally absent.\footnote{
This implies that the asymptotic completeness \cite{StreaterWightman},
which requires that the Heisenberg fields are expressed in terms of
the asymptotic fields, is lost if we restrict our consideration only
to the original fields.}

The absence of the pole in the Green's function $\VEV{\Psi\bPsi}$
remains in general true in our perturbation theory so long as
\SUNL$\otimes$\SUNR\ is a good quantum number.
The reason is as follows.
A possible origin of the discrete pole in $\VEV{\Psi\bPsi}$ is
one-particle intermediate propagation of the auxiliary field $\psi$
(see Fig.\ \ref{fig:pp}).\epsfhako{fig_pp.eps}{hbt}{8cm}{
A possible origin of the discrete pole in $\vev{\Psi\bPsi}$.
The solid line is the $\psi$ propagator, and the shaded blobs
represent complicated interactions.
}{fig:pp}{}
However, this is impossible since $\psi$ is \SUNR\ singlet while
$\Psi$ is \SUNR\ non-singlet.
As for the gauge field $A_\mu$, the situation is essentially the same
as the quark field explained above.
Note that, if the whole system including the PGM sector is treated in
naive perturbation theory, the two-point function (\ref{eq:PsiPsi})
has a discrete pole since we have
$g\sim 1 + \pi^a t^a$ with $\pi^a$ ($a=1,\ldots,N^2-1$) being the
Nambu-Goldstone mode.

Next, we shall look at the KO mechanism in the context of the
absence of the Nambu-Goldstone mode coupled to the BRST-exact color
current $N_\mu$ (\ref{eq:N}).
This is almost trivially satisfied in any finite order in our
perturbation expansion, since the \SUNL$\otimes$\SUNR\ symmetry
remains to  be realized in a linear manner and $N_\mu$ is nothing but
the Noether current of \SUNR.
However, we shall see how it is satisfied more concretely.
The following is almost the repetition of the argument for the quark
confinement given above.
The Nambu-Goldstone mode coupled to $N_\mu$, if it exists, should be
revealed as a massless pole in the two-point function
$\VEV{N_\mu {\cal O}}$ with ${\cal O}$ being an operator constructed
from the original fields, for example ${\cal O}=A_\nu$ \cite{KO}.
Within our perturbation theory the only possible origin of the
massless pole contributing to this Green's function is the auxiliary
gauge field $a_\mu$ with the free propagator (\ref{eq:Pa}).
However, the diagrams having a single $a_\mu$ as its
intermediate state vanish due to the conflict of the
\SUNL$\otimes$\SUNR\ quantum number ($a_\mu$ is \SUNR\ singlet while
$N_\mu$ and ${\cal O}$ are \SUNR\ non-singlets).

Summarizing, what is important for (and almost equivalent to) the KO
confinement mechanism is that the gauge-mode $g(x)$ is in the
disordered phase and the \SUNL$\otimes$\SUNR\ symmetry is realized in
a linear manner.
This property holds in the PGM and also in any finite order in our
perturbation expansion.
However, it is not clear at all whether this property persists beyond
our perturbation expansion in $\gYM$, and this is our most important
future subject.

\reseteqnum
\section{Finite temperature system}\label{sec:FT}

In this section we shall apply our perturbation theory to the finite
temperature case. In particular we are interested in what kind of PGM
is obtained as the $\gYM\to 0$ limit of finite temperature Yang-Mills
theory.
The partition function of the finite temperature Yang-Mills system is
given by
\begin{equation}
Z_{\rm FT}=\int_{\rm periodic}\D A_\mu\D c\D\bc\D B\exp\left\{
i\Int dx\Bigl[\Fmn(A)-i\dB G(A_\mu,c,\bc,B)\Bigr]\right\},
\label{eq:Z-ft}
\end{equation}
where all the path-integrals should be carried out using the
periodic boundary condition for the complex time
\cite{Bernard,FiniteT-HK}.
We have omitted the quark fields from the system for simplicity.
The following argument is applicable to both the imaginary and real
time formalisms although we should consider the real time one if we are
interested in the color (de)confinement at high temperature in the KO
mechanism \cite{Hata-Taniguchi}.

We shall generalize the Faddeev-Popov trick used in Sec.\
\ref{sec:perturbation} to the finite temperature case.
Eq.\ (\ref{eq:FPD}) for the FP determinant should now be
modified to
\begin{equation}
1=\Delta [A]\frac{1}{N}\sum_{k=0}^{N-1}
\int_{B_k}\D g\prod_{x}\delta\left(\p^\mu A_\mu^{g^{-1}}(x)\right),
\label{eq:FPD-ft}
\end{equation}
where $B_k$ denotes the boundary condition for $g(t,\bm{x})$ related
by the $Z_N$ twist:
\begin{eqnarray}
B_k:\quad
g(-i\beta,\bm{x}) = {\rm e}^{i\frac{2\pi k}{N}} g(0,\bm{x})\quad
(k=0,1,\cdots ,N-1).
\label{eq:BC}
\end{eqnarray}
This boundary condition is understood from the fact that $A_\mu^g$
should obey the same periodic boundary condition as $A_\mu$;
$A_\mu^g(-i\beta,\bm{x})=A_\mu^g(0,\bm{x})$.
The reason why we have to sum over all the boundary conditions
(\ref{eq:BC}) is that there does not necessarily exist $g(t,\bm{x})$
which satisfies the gauge condition $\p^\mu A_\mu^g=0$ for an
arbitrary $A_\mu$ if we restrict $g(t,\bm{x})$ to the periodic one.
Since the summation in (\ref{eq:FPD-ft}) is done with an equal weight,
the FP determinant is invariant, $\Delta[A^h]=\Delta[A]$,
under a gauge transformation $h$ with the twisted boundary condition;
$h(-i\beta,\bm{x})=\exp(2\pi i l/N)h(0,\bm{x})$.
Carrying out the same manipulation as in Sec.\ \ref{sec:perturbation},
we get the finite temperature generalization of Eq.\ (\ref{eq:ZF})
(we omit the source term):
\begin{eqnarray}
&& Z_{\rm FT} = \frac{1}{N}\sum_{k=0}^{N-1}
\int_\Bk\D g\int_{\rm periodic}\D c\D\bc\D B
\!\int_{\rm periodic}\D a_\mu\D\gamma\D\bgamma\D\beta\ne\\
&&\times \exp i\Int dx
\left[\Fmn(a) -i\edB H(a_\mu, \gamma, \bgamma, \beta)
-i\dB G\left(g^{\dagger}(a_\mu + \p_\mu)g, c, \bc, B\right)\right].
\label{eq:ZF-ft}
\end{eqnarray}
Eq.\ (\ref{eq:ZF-ft}) tells that the $\gYM\to 0$ limit of the finite
temperature Yang-Mills theory (without quark fields) is the PGM with
summation over the $Z_N$ boundary conditions:
\begin{eqnarray}
Z_{\mbox{\scriptsize FT-PGM}} = \frac{1}{N}\sum_{k=0}^{N-1}
\int_\Bk
\D g \int_{\rm periodic}\D c\D\bc\D B
\exp \left\{i\int dx \left [-i \dB G
\left(g^{\dagger}\p_\mu g, c, \bc, B\right)\right]\right\}.
\label{eq:PGM-ft}
\end{eqnarray}

The same formula as (\ref{eq:ZF-ft}) is also obtained using a
different method.
This is first to consider the partition function of a
topological system of the set of fields $(g,\gamma,\bgamma,\beta)$:
\begin{equation}
Y[A_\mu]=\sum_{k=0}^{N-1}
\int_\Bk\D g \int_{\rm periodic}\D \gamma \D \bgamma \D\beta
\exp\left( -\edB H[A_\mu^{g^{-1}},\gamma,\bgamma,\beta]\right).
\label{eq:Y}
\end{equation}
This $Y[A_\mu]$ is in fact independent of $A_\mu$. The invariance of
(\ref{eq:Y}) under an arbitrary infinitesimal deformation of
$A_\mu$, $A_\mu\to A_\mu + \delta A_\mu$, is a consequence of the
$\edB$-exactness of the action and the $\edB$ invariance of the
path-integral measure. Owing to the summation over all the
boundary conditions (\ref{eq:BC}) with an equal weight, $Y[A_\mu]$ is
invariant under the ``large'' gauge transformation, $A_\mu\to A_\mu^h$
with $h(-i\beta,\bm{x})=\exp(2\pi i l/N)h(0,\bm{x})$.
Since (\ref{eq:Y}) is independent of $A_\mu$, we insert (\ref{eq:Y})
into Eq.\ (\ref{eq:Z-ft}), making the substitution
(\ref{eq:expansion}), and we finally get Eq.\ (\ref{eq:ZF-ft}).

The finite temperature PGM was discussed in Ref.\
\cite{Hata-Taniguchi} and the non-perturbative deconfining dynamics
was analyzed there.  Our finding was that in the low
temperature region $g$ is in the disordered (confining) phase,
however, in the high temperature region the ordered (deconfining)
phase is realized. The deconfining transition occurred since the
infrared singularity in the perturbative phase is softened in the
twisted boundary condition sectors.
In Ref.\ \cite{Hata-Taniguchi} we summed over more general boundary
conditions than (\ref{eq:BC});
$g(-i\beta,\bm{x})=h\!\cdot\! g(0,\bm{x})$ with $h$ being a general
element of the gauge group SU$(N)$.
However, the essential properties of the deconfining transition
observed in Ref.\ \cite{Hata-Taniguchi} remain unchanged even if the
boundary conditions are restricted to the $Z_N$ ones as in
(\ref{eq:PGM-ft}).

If we introduce the quark field $\Psi$ (which is subject to the
anti-periodic boundary condition) into the system, the auxiliary quark
field $\psi=g\Psi$ in the $B_k$ sector should obey the boundary condition
$\psi(-i\beta,\bm{x})=-\exp\left(2\pi ik/N\right)\psi(0,\bm{x})$.
The (de)confinement of the original quark field $\Psi$ is shown
similarly to the zero-temperature case of Sec.\ \ref{sec:KO}.

\reseteqnum
\section{Discussion}\label{sec:discussion}

In this paper we have presented a perturbation expansion to
incorporate the physical degrees of freedom into the PGM.
We have seen that the quarks and gluons are confined in this
perturbation theory.

Of course this is not the end of the confinement problem.
We have to know whether the confining property persists in a more
complete analysis of QCD including the non-perturbative treatment in
$\gYM$.
In particular, we are interested in whether the disordered phase of
the gauge-mode $g(x)$ remains intact beyond perturbation
theory.\footnote{
As good news for confinement, we mention
the one-loop beta function of the PGM coupling constant $\lambda$,
$\beta_\lambda = d\lambda/d\ln\mu =-\lambda^2 -3\gYM^2\lambda$
\cite{HataNiigata}.
This $\beta_\lambda$ tells that $\gYM$ helps to strengthen the infrared
slavery of $\lambda$.}
For studying non-perturbative effects to the dynamics of the
gauge-mode, it may be helpful to consider the ``effective action''
$S_{\rm eff}$ by regarding (\ref{eq:Zpert}) with $J=0$ as a modified
topological model:
\begin{equation}
S_{\rm eff}(g,c,\bc,B)= -i\Int dx\dB G(g^\dagger\p_\mu g,c,\bc,B)
+ W\left[-\frac{4i}{\lambda}\dA\dB\left(g\p_\mu
g^\dagger\right)\right].
\label{eq:Seff}
\end{equation}
Note that the $W$ term in (\ref{eq:Seff}) is also written in a
BRST-exact form $\dB(\cdots)$.

As a related question, we have not yet understood how the confinement
breaks down in the Higgs phase. Namely, if we introduce the Higgs
scalar into the system and break the color symmetry spontaneously, it
is natural to expect that the color confinement fails.
However, in the lowest order in our perturbation theory, the color
confinement holds irrespectively of the presence of the Higgs field.
This problem may be understood on the basis of the effective topological
model (\ref{eq:Seff}).
In the case the color gauge symmetry is broken in an asymmetric
manner and the gauge bosons get different masses, this effect appears
in $W$ of (\ref{eq:Seff}) as explicit breaking terms of the \SUNL\
symmetry as seen from the lowest order expression (\ref{eq:Wfirst}).
This explicit breaking may trigger the ordered phase of the
gauge-mode $g(x)$ to break the confinement.
However, if the gauge bosons acquire an equal mass (for example, the
case the SU$(2)$ gauge symmetry is broken by a doublet Higgs field),
the $W$ term of (\ref{eq:Seff}) has no explicit breaking of
\SUNL$\otimes$\SUNR.
The understanding of the failure of confinement in this case is a
challenging problem in the future work.

\vspace{.5cm}
\reseteqnum
\renewcommand{\thesection}{A}
\noindent
{\Large\bf Appendix: BRST transformations}
\vspace{.5cm}

Here, we summarize the (anti-)BRST transformations, $\dB$, $\dA$,
$\edB$ and $\edA$. First, $\dB$ and $\dA$ are defined by
\begin{eqnarray}
\dB A_\mu=\CD_\mu(A)c,\ \
\dB g=gc,\ \
\dB c=-c^2,\ \
\dB\bc=iB,\ \
\dB B=0,\ \
\dB\Psi=c\Psi,\\
\dA A_\mu=-\CD_\mu(A)\bc,\ \
\dA g=-g\bc,\ \
\dA\bc=\bc^2,\ \
\dA c=-i\bB,\ \
\dA\bB=0,\ \
\dA\Psi=-\bc\Psi,
\end{eqnarray}
with $B+\bB=i\{c,\bc\}$.
The auxiliary fields \auxfields\ are inert under $\dB$ and $\dA$.
Next are $\edB$ and $\edA$:
\begin{eqnarray}
\edB a_\mu=\CD_\mu(a)\gamma,\ \
\edB g=-\gamma g,\ \
\edB\gamma=-\gamma^2,\ \
\edB\bgamma=i\beta,\ \
\edB\beta=0,\ \
\edB\psi=\gamma\psi,
\\
\edA a_\mu=-\CD_\mu(a)\bgamma,\ \
\edA g=\bgamma g,\ \
\edA\bgamma=\bgamma^2,\ \
\edA\bgamma=-i\bbeta,\ \
\edA\bbeta=0,\ \
\edA\psi=-\bgamma\psi,
\end{eqnarray}
where $\beta+\bbeta=i\{\gamma,\bgamma\}$.
The original fields $(A_\mu,c,\bc,B,\Psi)$ are invariant under
$\edB$ and $\edA$.

\newcommand{\J}[4]{{\sl #1} {\bf #2} (19#3) #4}
\newcommand{\MPL}{Mod.\ Phys.\ Lett.} \newcommand{\NP}{Nucl.\ Phys.}
\newcommand{\PL}{Phys.\ Lett.} \newcommand{\PR}{Phys.\ Rev.}
\newcommand{\PRL}{Phys.\ Rev.\ Lett.} \newcommand{\AP}{Ann.\ Phys.}
\newcommand{\CMP}{Commun.\ Math.\ Phys.}
\newcommand{\PTP}{Prog.\ Theor.\ Phys.}

\end{document}